\documentclass[11pt, draftclsnofoot, onecolumn]{IEEEtran}

\usepackage[utf8]{inputenc} 
\usepackage[T1]{fontenc}
\usepackage{url}              
\usepackage{cite}             

\usepackage[cmex10]{amsmath}  
\usepackage{mleftright}       
\mleftright                   

\usepackage{graphicx}         
\usepackage{booktabs}         

\usepackage[pass]{geometry} 

\usepackage{cite}
\usepackage{pgfplots}
\usepackage{csquotes}
\usepackage{amsmath,amssymb,amsfonts,amsthm}
\usepackage{algorithmic}
\usepackage{graphicx}
\usepackage{textcomp}
\usepackage{xcolor}
\usepackage{xcolor, soul}
\sethlcolor{cyan}
\def\BibTeX{{\rm B\kern-.05em{\sc i\kern-.025em b}\kern-.08em
		T\kern-.1667em\lower.7ex\hbox{E}\kern-.125emX}}
\usepackage{bm}
\usepackage{enumitem}       
\usepackage{url}            
\usepackage{booktabs}       
\usepackage{caption}
\usepackage{tikz}
\usepackage[lined,boxed,linesnumbered,ruled,vlined]{algorithm2e}
\usepackage{comment}
\usepackage{balance}
\usepackage[capitalize]{cleveref} 
	\crefname{equation}{}{}
	\crefname{theorem}{Theorem}{Theorems}
	\crefname{lemma}{Lemma}{Lemmas}
	\crefname{cor}{Corollary}{Corollaries}
	\crefname{prop}{Proposition}{Propositions}
	\crefname{note}{Note}{Notes}
	\crefname{appsec}{Appendix}{Appendices}
	\crefname{definition}{Definition}{Definitions}
	\crefname{conj}{Conjecture}{Conjectures}
	\crefname{construction}{Construction}{Constructions}
\usepackage{ifthen}
\newboolean{IncludeRandomWalk} 
\setboolean{IncludeRandomWalk}{false}
\newboolean{DecreasingLearningRate} 
\setboolean{DecreasingLearningRate}{false} 
\usepackage{nicefrac}  
\usepackage{mathrsfs}  
\usepackage{float}
\usepackage[caption=false,font=footnotesize,labelformat=simple]{subfig}

\usetikzlibrary{shapes, patterns,decorations.text, decorations.pathreplacing}

\newcommand{\set}[1]{\mathcal{#1}}

\newcommand{\EE}[2][]{
	\ifthenelse{\equal{#1}{}}%
	{\mathbb{E} \left[ #2 \right]}
	{\mathbb{E}_{#1} \left[ #2 \right]}
}

\newtheorem{theorem}{Theorem}

\newtheorem{lemma}{Lemma}

\newtheorem{proposition}{Proposition}
\allowdisplaybreaks

\usepackage[utf8]{inputenc} 
\usepackage[T1]{fontenc}
\usepackage{url}
\usepackage{ifthen}
\usepackage{cite}
\usepackage{xcolor}
\usepackage[cmex10]{amsmath} 
\usepackage{graphicx}
\usepackage{amsmath}
\usepackage{flushend}
\usepackage{pgfplots}
\pgfplotsset{compat=1.17}

\begin{document}

\title{Context-Aware Search and Retrieval Over\\ Erasure Channels}

\author{%
  \IEEEauthorblockN{Sara Ghasvarianjahromi, Yauhen Yakimenka, J\"org Kliewer}
  \IEEEauthorblockA{Helen and John C. Hartmann Department of Electrical and Computer Engineering\\ New Jersey Institute of Technology, Newark, New Jersey, 07102, USA
\\Email: \{sg273, yauhen.yakimenka, jkliewer\}@njit.edu
}

 }

\maketitle

\begin{abstract}
  This paper introduces and analyzes a search and retrieval model that adopts key semantic communication principles from retrieval-augmented generation. We specifically present an information-theoretic analysis of a remote document retrieval system operating over a symbol erasure channel. The proposed model encodes the feature vector of a query, derived from term-frequency weights of a language corpus by using a repetition code with an adaptive rate dependent on the contextual importance of the terms. At the decoder, we select between two documents based on the contextual closeness of the recovered query. By leveraging a jointly Gaussian approximation for both the true and reconstructed similarity scores, we derive an explicit expression for the retrieval error probability, i.e., the probability under which the less similar document is selected.
 Numerical simulations on synthetic and real-world data (Google NQ) confirm the validity of the analysis.
They further demonstrate that assigning greater redundancy to critical features effectively reduces the error rate, highlighting the effectiveness of semantic-aware feature encoding in error-prone communication settings.  
\end{abstract}

\section{Introduction}
 In recent years, search and retrieval systems have been widely applied across various domains, including search engines, question-answering frameworks, and recommender systems \cite{buttcher2016information, siriwardhana2023improving}. These systems are fundamental to organizing and accessing the vast amounts of information generated daily, enabling users to efficiently locate relevant data and insights. For instance, question-answering systems have gained significant attention as they aim to go beyond document retrieval by providing precise and context-aware answers to user queries, rather than merely returning a list of relevant documents \cite{dimitrakis2020survey, zhu2021retrieving, lan2021semantic}.

A related field that has experienced significant growth is semantic communication, which prioritizes preserving the meaning and context of transmitted information over achieving perfect bit-level reconstruction \cite{shi2021semantic, lu2023semantics,luo2022semantic}. 
Early studies in semantic communication focused on meaningful representations that reduce overhead while preserving the semantics \cite{lan2021semantic}.
Subsequent works integrated machine learning and neural networks to build end-to-end frameworks, where encoders and decoders capture and reconstruct essential features \cite{yang2022semantic, chaccour2024less}. 
A notable contribution to this domain is the work on importance-aware power allocation guided by pre-trained language models (e.g., BERT or  GPT) for semantic communication \cite{guo2023semantic}.

Another notable development at the intersection of information retrieval and natural language processing is retrieval-augmented generation (RAG). RAG systems combine traditional information retrieval techniques, often leveraging TF-IDF, BM25, or pre-trained embeddings with generative models to produce coherent and contextually rich outputs \cite{gao2023retrieval, salemi2024evaluating}. They typically store document embeddings in a database and retrieve the most relevant ones by measuring query–document similarity using metrics like cosine distance or L2 norms \cite{li2024enhancing, bhattarai2024enhancing, 10603291}. Recently, large language models have further enhanced RAG architectures by providing improved feature extraction and more accurate similarity assessments \cite{soong2024improving}. 
However, to the best of our knowledge, RAG systems have not yet been explored or analyzed from an information-theoretic perspective. 
This specifically holds true under the additional assumption of erroneous feature vectors, where components could be erased by communication imperfections or the action of an adversary. We aim to fill this void in this paper.

Building on these two directions, this paper introduces and analyzes a simplified model that adopts key semantic communication principles within a RAG-like system. Our proposed model incorporates fundamental concepts such as computing word weights as features and retrieving relevant documents through similarity assessments between the query and the documents. Additionally, our model addresses key challenges in error-prone communication environments, including the reliable transmission of contextual query features over a symbol erasure channel, resilience to channel impairments, and accurate document retrieval at the receiver. We show that the importance of specific features governs the addition of redundancy in such a way that the context and the semantic meaning are preserved. 
For the sake of a tractable analysis we focus on a simplified setting for only two documents, TF-IDF for feature extraction, simple repetition coding, and a squared L2-norm as the similarity measure.

Specifically, we develop an analytical framework to characterize the error probability  in retrieving the contextually closest document, when the user query features are encoded by a repetition code and then passed through a symbol erasure channel.
We show that both the true and the reconstructed similarity metrics are asymptotically jointly Gaussian, which enables us to derive a closed-form expression for the retrieval error probability.
 Our results confirm that, verified by numerical simulations,  encoding more important terms with higher redundancy effectively lowers the probability of error, in the presence of query feature erasures.

\section{System model}
\label{sec:system_model}
\begin{figure*}[t]
    \centering
    \includegraphics[width=\textwidth]{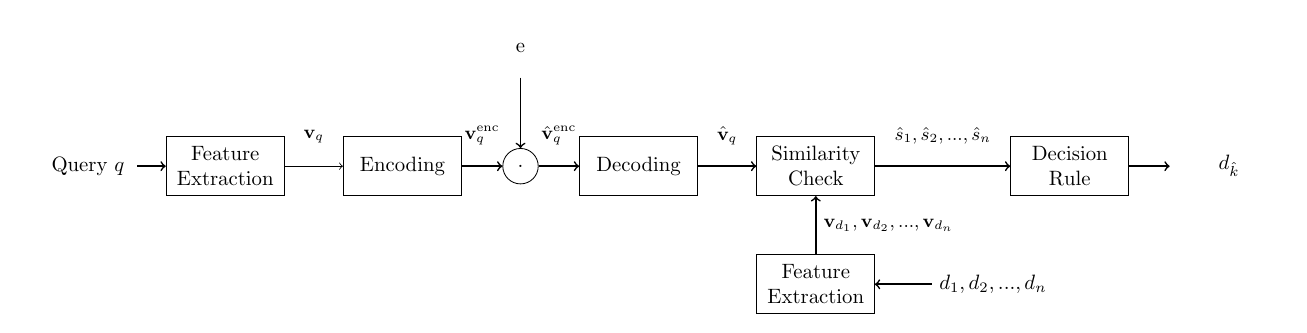}
    \vspace{-10pt}
    \caption{System model.}
    \label{fig:sm}
\end{figure*}

We consider a remote document retrieval system designed to identify the document most relevant to a given query $q$ consisting of $l$ terms denoted as $q=\{w_1,w_2,\dots,w_l\}$.
By ``terms'' we mean individual words or meaningful components of the query text.
Every term in the query $q$ is an element of the predefined vocabulary $\mathcal V=\{t_1, t_2, \cdots, t_N\}$ consisting of $N$ distinct terms, sorted by their ranks, i.e., the position in a list of words ordered by their frequency of occurrence.
For example, the vocabulary can be all the words of the English language. 
The system consists of a transmitter, a receiver, and an erasure channel, as illustrated in Fig. \ref{fig:sm}.
The query is transformed into a feature vector $\mathbf v_q\in \mathbb R^N$ at the transmitter side. This vector resides in an $N$-dimensional real space with at most $l$ non-zero elements, where each dimension corresponds to a term in $\mathcal V$.
Therefore, a ``feature vector'' is a numerical representation that quantifies the importance of these terms within the query. 
This feature vector is then encoded to counter the potential loss of information caused by the erasure channel.

The encoding process begins by forming index-value pairs $(i,v_{q,i})$ for all non-zero elements of the feature vector.
Here, $i\in[N]$ is the index corresponding to a term, and $v_{q,i}$ is its associated importance weight. 
To ensure robustness against erasures, a repetition code is applied, where each pair is repeated $r_i$ times.
This repetition introduces redundancy, forming the encoded $\mathbf v_q^{\mathrm{enc}}$, which is then transmitted through the erasure channel.
The erasure channel is a communication medium where transmitted symbols (i.e., $(i,v_{q,i})$ pair) may be erased with a certain probability $\epsilon$.
In this context, ``erased'' means that the receiver does not receive some of the transmitted symbols, leading to information loss.
The goal of introducing redundancy during encoding is to enable accurate reconstruction of the original feature vector even in the presence of erasures.

At the receiver side, the transmitted message $\mathbf{\hat{v}}_q^{\mathrm{enc}}$ is retrieved and decoded into the reconstructed feature vector $\mathbf{\hat{v}}_q\in\mathbb{R}^N$.
The receiver then performs a similarity check by comparing $\mathbf{\hat{v}}_q$ with the feature vectors of all documents in the corpus $\mathcal D$, represented as $\{\mathbf v_{d_1}, \mathbf v_{d_2},\dots, \mathbf v_{d_n}\}\in\mathbb{R}^N$, corresponding to documents $\{d_1,d_2,\cdots,d_n\}$ .
The ``corpus'' refers to the collection of documents stored in the system.
The similarity scores $\hat{s}_1,\hat{s}_2,\dots,\hat{s}_n$ between $\mathbf{\hat{v}}_q$ and the document feature vectors are calculated using a predefined similarity measure, i.e., L2-norm squared. 
Based on these scores, the document $d_{\hat{k}}$ with the closest similarity is identified as the one most relevant to the query. 
This decision is made using a predefined decision rule. 
\vspace{-20pt}

\section{Analysis}
\label{sec:analysis}
In this section, we present a detailed analysis  of each component in the system model, shown in Fig. \ref{fig:sm}.
\subsection{Vocabulary and Query}
As described in the system model, we assumed a predefined vocabulary $\mathcal V$ with $N$ distinct terms.
The frequencies of the terms within this vocabulary follow Zipf's law \cite{piantadosi2014zipf}.
According to this law, each term $t_i\in\mathcal V$ is assigned a unique rank $i$ based on its frequency in the vocabulary, with $t_1$ being the most frequent term and $t_N$ being the least frequent.
The frequency $f_i\propto \nicefrac{1}{i^\alpha}$ of a term is inversely proportional to its rank $i$, where $\alpha\geq0$  is the Zipfian exponent that controls the decay rate of the term frequencies. 
This relationship characterizes the heavy-tailed distribution observed in a natural language, where a few terms occur very frequently, and the majority appear rarely. 
To transform this relationship into a probability distribution, we introduce a normalization factor $\sum_{j=1}^N{{1}/{j^\alpha}}$.
The probability of selecting a term at rank $i$, in accordance with Zipf's law, is given by
\vspace{-5pt}
\begin{equation}
    p_i = \frac{{1}/{i^\alpha}}{\sum_{j=1}^N{{1}/{j^\alpha}}}.
    \label{eq:zipfs_distribution}
\end{equation}
To model a query $q=\{w_1,w_2,\dots,w_l\}$, we assume that its terms are drawn independently from the vocabulary $\mathcal V$ following Zipf's distribution.
Consequently, vector $\mathbf c_q$ with length $N$, represents the count of each term in the query, and the multinomial distribution models the frequency with which each term appears in a query:
\vspace{-3pt}
\begin{equation}
    \mathbf c_{q}\sim\mathrm{Multinomial}(l, \{p_1, p_2, \cdots, p_N\}).
\end{equation}
Here $\{p_1,p_2, \cdots,p_N\}$ represent the probabilities of the terms in the vocabulary, determined by their ranks in accordance with Zipf's law \cref{eq:zipfs_distribution}. 
The mean and covariance matrix of $\mathbf c_q$ are given as \cite{wasserman2013all}
\vspace{-5pt}
\begin{equation}
    \mathbf{\mu}_{\mathbf{c}_q}=
    \begin{pmatrix}
        l p_1 \\
        \vdots \\
        l p_N
    \end{pmatrix}=l{\mathbf p},
\end{equation}
and 
\vspace{-5pt}
\begin{equation}
    \mathbf{\Sigma}_{\mathbf c_q} =
\begin{pmatrix}
    l p_1 (1 - p_1) & -l p_1 p_2 & \cdots & -l p_1 p_N \\
    -l p_1 p_2 & l p_2 (1 - p_2) & \cdots & -l p_2 p_N \\
    \vdots & \vdots & \ddots & \vdots \\
    -l p_1 p_N & -l p_2 p_N & \cdots & l p_N (1 - p_N)
\end{pmatrix}.
\end{equation}
\subsection{Feature Extraction}
To obtain the relative frequencies of terms in the query, we normalize the count vector $\mathbf c_q$ by dividing the count of each term by the length of the query $l$, so that $\Tilde{\mathbf{c}}_q= \frac{\mathbf{c}_q}{l}$ ensures the property $\sum_{i=1}^N{\Tilde{{c}}_{q,i}=1}$. 
This approach aligns directly with the term frequency (TF) method commonly used in natural language processing and information retrieval, where the term frequency for a term $i$ in a query is defined as \cite{mishra2015analysis, ibrihich2022review}
\begin{equation}
    \mathrm{TF}(i) = \frac{\text{Number of occurrences of } \text{term } i \text{ in the query}}{\text{Total number of terms in the query (} l \text{)}}.
\end{equation}
Here, $\Tilde{{c}}_{q,i}$ corresponds to $\mathrm{TF}(i)$.
For large $l$, the normalized vector $\Tilde{\mathbf{c}}_q$ has approximately a multivariate normal distribution \cite[Thrm. 14.5]{wasserman2013all} as
\vspace{-4pt}
\begin{equation}
    \Tilde{\mathbf{c}}_q\sim\mathcal{N}(\mathbf{\mu}_{\Tilde c_q},\mathbf{\Sigma}_{\Tilde c_q}),
\end{equation}
where $\mathbf{\mu}_{\Tilde{\mathbf{c}}_q}=\frac{1}{l}\mathbf{\mu}_{{\mathbf{c}}_q}$ is the mean vector, and $\mathbf{\Sigma}_{\Tilde{\mathbf{c}}_q}=\frac{1}{l^2}\mathbf \Sigma_{{\mathbf{c}}_q}$ is the covariance matrix, capturing the variance and covariances of the term frequencies.

\subsection{Encoding}
To address the challenges posed by the erasure channel and ensure reliable transmission of important terms, we first derive a modified feature vector $\mathbf v_q$ from the normalized count vector $\mathbf{\Tilde c}_q$.
Specifically, we cut out a set of length $l_s$ of frequent but low-informative terms (referred to  as ``stop words''  in NLP \cite{fox1989stop}) from $\mathbf{\Tilde c}_q$ by forcing those coordinates to $0$.
Concretely, define a linear operator $\mathcal G$ that acts on an $N$-dimensional vector by:
\vspace{-3pt}
\begin{equation}
    \mathcal G(x)_i=\begin{cases}
0, & \text{if $i$ is one of the stop-word indices},\\
x_i, & \text{otherwise}.
\end{cases}
\end{equation}
Then, the modified vector $\mathbf v_q$ is $\mathcal G\left(\mathbf{\Tilde{\mathbf{c}}}_q\right)$.
In this case, $\mathbf v_q$ is still a linear transformation of a Gaussian vector, so it remains Gaussian with mean $\mathbf{\mu}=\mathcal G\mathbf{\mu}_{\mathbf{\Tilde{c}}_q}$ and covariance matrix $\mathbf{\Sigma}=\mathcal G\mathbf{\Sigma}_{\mathbf{\Tilde{c}}_q}\mathcal G^T$.
This modification is motivated by the fact that the encoding of nonzero elements directly depends on each term's TF-score, reflecting its importance. 
By eliminating those most frequent terms, we reduce the coding overhead without significantly affecting the semantic content, thereby preserving retrieval accuracy.

Although $\mathbf{\Tilde{c}_q}$ itself sums to 1, the sum of the entries of $\mathbf v_q$ may be less than 1 due to the applied modification.
Let $M$ be the number of nonzero entries in $\mathbf v_q$ after modification.
The encoding process begins by forming an index-value pair $(i,v_{q,i})$ for each nonzero element ($v_{q,i}\neq0$) in $\mathbf v_q$.
This avoids the need to transmit the entire vector, focusing instead on the nonzero elements along with their indices.
We then apply a repetition code to all ($i, v_{q,i}$) pairs.
To control the transmission rate, we define a code rate $R=\nicefrac{M}{\sum_{i=1}^N{r_i}}$, where $r_i$ denotes the number of repetitions allocated to the $i$-th nonzero term.
Because the number of repetitions should be proportional to each term’s TF-based importance, we set
\vspace{-2pt}
\begin{equation}
r_i = \left\lceil \frac{M}{R\sum_{i=1}^N{v_{q,i}}} {v}_{q,i} \right\rceil.
\label{eq:repetition}
\end{equation}
\subsection{Erasure Channel}
We model transmission over an erasure channel with erasure probability $\epsilon$.
For each transmitted symbol $v_{q,i}^{\rm{enc}}$, we assume that each repetition is erased independently with probability $\epsilon$.
Consequently, if a pair $(i,v_{q,i})$ is repeated $r_i$ times, the probability that at least one copy is successfully received is $1-\epsilon^{r_i}$, and the probability that all copies are erased is $\epsilon^{r_i}$.
To formalize this behavior, we define the erasure pattern $\mathbf e=[e_i]$ with 
\vspace{-3pt}
\begin{equation}
e_i =
\begin{cases} 
1, & \text{with probability } 1 - \epsilon^{r_i}, \\
0, & \text{with probability } \epsilon^{r_i}.
\end{cases}
\label{eq: rep_effect}
\end{equation}
This notation $\mathbf e=[e_i]$ allows us to model the erasure channel behavior probabilistically, where $e_i=1$ indicates successful reception of the $i$-th symbol, and $e_i=0$ indicates erasure.
The overall outcome for the $i$-th pair, after transmission through the erasure channel, can then be described as:
\vspace{-3pt}
\begin{equation}
\hat{{v}}_{q,i}^{enc} =
\begin{cases} 
{v}_{q,i}^{enc}, & \text{if}\,\,\, e_i=1, \\
?, & \text{if}\,\,\, e_i=0.
\end{cases}
\end{equation}
where ``$?$'' indicates that no copy of $(i,v_{q,i})$ was received.
\subsection{Decoding}
At the receiver side, we initialize the reconstructed vector $\hat{\mathbf v}_q$ as a zero vector of length $N$, where $N$ is the known vocabulary size. 
For each transmitted index-value pair $(i,v_{q,i})$, repeated $r_i$ times over the erasure channel, if at least one copy arrives successfully, we set the $i$-th component of $\hat{\mathbf{v}}_q$ to ${v}_{q, i}$. 
Otherwise, if all copies are erased (or if $v_{q,i}$ was originally zero), $\hat{{v}}_{q,i}$ remains zero, i.e.,  $\hat{\mathbf v}_q=\mathbf v_q\circ \mathbf e$, where \(\circ\) denotes the Hadamard product.
\subsection{Similarity Check and Decision Rule}
At the receiver side, the goal is to identify the document $d_{\hat{k}}$ from the corpus $\mathcal D$  containing $n$ documents $\{d_1,d_2,\cdots,d_n\}$, that is most relevant to the query $q$. 
We assume that the set of all unique terms appearing in the corpus is contained within the predefined vocabulary $\mathcal{V}$, as $\bigcup_{j=1}^{n} \mathcal{T}_{d_j} \;\subseteq\; \mathcal{V}$, where $\set T_{d_j} = \{\,t \mid t \in d_j\}$ represents the set of terms in document $d_j$.
To compare each document with the query, the receiver computes the feature vectors, namely the TF-scores for all documents in the corpus, denoted as $\mathbf{v}_{d_1}, \mathbf{v}_{d_2}, \dots, \mathbf{v}_{d_n}$. It then calculates the inverse document frequency (IDF) score for each term in the vocabulary, $t_i \in \mathcal{V}$. The IDF score quantifies the importance of a term across the corpus, emphasizing terms that are rare.
Formally, the IDF score is \cite{zhao2018tfidf}
\vspace{-2pt}
\begin{equation}
  \mathrm{IDF}(i) 
  \;=\; \log\!\Bigl(\frac{n+1}{n_{i} + 1}\Bigr),
  \end{equation}
where $n_{i}$ is the number of documents that contain the $i$-th term. Let $\xi\in\mathbb R^N$ be the vector of these IDF scores.

Next, the receiver evaluates how close the reconstructed query vector $\hat{\mathbf{v}}_q$ is to each document vector $\mathbf{v}_{d_j}$ in the TF-IDF space. Specifically, it computes
\vspace{-2pt}
\begin{equation}
  \hat{s}_j 
  \;=\; 
  \bigl\|\bigl(\hat{\mathbf{v}}_q - \mathbf{v}_{d_j}\bigr) \,\circ\, \xi\bigr\|^2,
  \quad j = \{1, 2, \dots, n\},
\end{equation}
where $\|\cdot\|^2$ represents the (squared) L2-norm. 
This metric leverages the contextual dependencies inherent in language by incorporating the importance of terms, similar to RAG systems.
The document with the smallest value of $\hat{s}_j$ is identified as the most relevant to the query:
\vspace{-2pt}
\begin{equation}
  \hat{k} 
  \;=\; 
  \arg\min_{\,j \in \{1, 2, \dots, n\}} \hat{s}_j.
\end{equation}
Hence, $d_{\hat{k}}$ is selected as the document most relevant to the query based on the minimum L2-norm    criterion.

\section{Results}
\label{sec:results}
In this section, we present both analytical and numerical results to evaluate the system's performance.
\subsection{Analytical Results}
\label{Theo: error_pr}
Our objective is to characterize the \emph{probability of error}, i.e., the probability that the receiver's decision $d_{\hat{k}}$ differs from the ground-truth document $d_{k}$ (which would be chosen in the absence of channel erasures). 
To facilitate the analysis, we focus on a corpus containing only two documents, i.e., $n = 2$.
Moreover, for the theoretical derivations in this section, we assume a fixed repetition counts $r_i=r$ to facilitate the analysis.
Note that the effect of repetition is captured by the erasure pattern $\mathbf e$ in (\ref{eq: rep_effect}).
Let
\vspace{-2pt}
\begin{equation}
    s_1 = \bigl\|\bigl(\mathbf{v}_q - \mathbf{v}_{d_1}\bigr) \circ \xi\bigr\|^2
\quad\text{and}\quad
s_2 = \bigl\|\bigl(\mathbf{v}_q - \mathbf{v}_{d_2}\bigr) \circ \xi\bigr\|^2,
\end{equation}
be the squared-error scores for documents $d_1$ and $d_2$ with respect to the query feature vector $\mathbf{v}_q$. Define $s = s_1 - s_2$. 
By expanding each squared norm and subtracting, we can show that $s$ is a linear transformation of $\mathbf{v}_q$:
\vspace{-5pt}
\begin{equation}
s \;=\; \sum_{i=1}^N \bigl[a_i\, {v}_{q,i} + c_i\bigr]
\;=\;
\mathbf{a}^T\,\mathbf{v}_q \;+\; \sum_{i=1}^{N}{c_i},
\label{eq:s_expansion}
\end{equation}
where $a_i \;=\; 2\,\xi_i^2\bigl(\mathbf{v}_{d_2,i} - \mathbf{v}_{d_2,i}\bigr)$ and $c_i \;=\; \xi_i^2\bigl(\mathbf{v}_{d_1,i}^2 - \mathbf{v}_{d_2,i}^2\bigr)$.
Therefore $s$ is asymptotically Gaussian, i.e., $s\sim\mathcal N(\mu_s,\sigma_s^2)$, and the Gaussian assumption is well justified for large vocabularies.
Hence, the mean and variance of $s$ are
\vspace{-6pt}
\begin{equation}
\mu_s 
\;=\;
\mathbb{E}[s] 
\;=\; 
\mathbf{a}^T\,\mathbf{\mu} \;+\; \sum_{i=1}^{N}{c_i},
\end{equation}
\vspace{-6pt}
\begin{equation}
\sigma_s^2
\;=\;
\mathrm{Var}(s)
\;=\;
\mathbf{a}^T\,\Sigma\,\mathbf{a},
\end{equation}
where $\mathbf{\mu} = \mathbb{E}[\mathbf{v}_q]$, $\Sigma = \mathrm{Cov}(\mathbf{v}_q)$, and $\mathbf a=[a_i]$.

Similarly, at the receiver side, the corresponding quantities become
\vspace{-5pt}
\begin{equation}
\hat{s}_1 
= 
\bigl\|\bigl(\hat{\mathbf{v}}_q - \mathbf{v}_{d_1}\bigr)\circ\xi\bigr\|^2,
\quad
\hat{s}_2 
=
\bigl\|\bigl(\hat{\mathbf{v}}_q - \mathbf{v}_{d_2}\bigr)\circ\xi\bigr\|^2,
\end{equation}
and $\hat{s}\;=\;\hat{s}_1 - \hat{s}_2$.

By analyzing the distributions of $s$ and $\hat{s}$, we can derive conditions under which the receiver’s choice $d_{\hat{k}}$ diverges from the true best document $d_{k}$, thereby determining the probability of error.
\begin{lemma}\label{lem:hat-s-gaussian}
    Consider a fixed erasure pattern $\mathbf{e}$, where $\hat{\mathbf{v}}_q = \mathbf{v}_q \circ \mathbf{e}$ (or equivalently $\hat{\mathbf{v}}_q = \mathbf{D}_e \,\mathbf{v}_q$). 
    Then the similarity difference $\hat{s} \mid {\mathbf{e}}$ asymptotically follows a normal distribution with mean $\mu_{\hat{s}\mid {\mathbf{e}}}$ and variance $\sigma^2_{\hat{s}\mid {\mathbf{e}}}$, i.e.,
    \begin{IEEEeqnarray*}{rcl}
      \hat{s}\mid {\mathbf{e}} \;\sim\; \mathcal{N}\!\bigl(\mu_{\hat{s}\mid {\mathbf{e}}}, \,\sigma_{\hat{s}\mid {\mathbf{e}}}^2\bigr).
    \end{IEEEeqnarray*}
\end{lemma}
\begin{IEEEproof}
    Recall that $\hat{s} \!=\! \hat{s}_1\! -\! \hat{s}_2$, where $\hat{s}_j \!=\! \|\bigl(\hat{\mathbf{v}}_q\! -\! \mathbf{v}_{d_j}\bigr)\circ \xi\|^2$.
    From \cref{eq:s_expansion} we know that $\hat{s}$ can be written similarly as 
    \vspace{-5pt}
    \begin{IEEEeqnarray*}{rcl}
      \hat{s}\mid {\mathbf{e}} \;=\; \mathbf{a}^T\,(\mathbf{v}_q \circ \mathbf{e}) \;+\; \sum_{i=1}^{N}{c_i}
      \;=\;
      \mathbf{a}^T\,\mathbf{D}_e\,\mathbf{v}_q \;+\;\sum_{i=1}^{N}{c_i},
    \end{IEEEeqnarray*}
    where $\mathbf{D}_e = \mathrm{diag}(\mathbf{e})$. 
    Since $\mathbf{v}_q$ is asymptotically Gaussian with mean $\boldsymbol{\mu}$ and covariance $\boldsymbol{\Sigma}$, any linear transformation of $\mathbf{v}_q$ remains Gaussian. Hence, $\hat{s}\mid {\mathbf{e}}$ is asymptotically Gaussian with
    \vspace{-10pt}
    \begin{IEEEeqnarray*}{rcl}
      \mu_{\hat{s}\mid \mathbf{e}}
      \;=\;
      \mathbf{a}^T \mathbf{D}_e \,\boldsymbol{\mu}
      \;+\;
      \sum_{i=1}^{N}{c_i},
      \quad\text{and}\quad
      \sigma_{\hat{s}\mid \mathbf{e}}^2
      \;=\;
      \mathbf{a}^T \mathbf{D}_e\,\boldsymbol{\Sigma}\,\mathbf{D}_e\,\mathbf{a}.
      \vspace{-2ex}
    \end{IEEEeqnarray*}
\mbox{}
\end{IEEEproof}
\begin{proposition}\label{pro:jointly_gaussian}
    The pair $\bigl(s,\,\hat{s}\mid {\mathbf{e}}\bigr)$ is asymptotically jointly Gaussian with mean vector
    $\bigl(\mu_s,\;\mu_{\hat{s}\mid {\mathbf{e}}}\bigr)^T$
    and covariance matrix
    \begin{IEEEeqnarray*}{rcl}
    \begin{pmatrix}
      \sigma_s^2 & \rho_e\,\sigma_s\,\sigma_{\hat{s}\mid {\mathbf{e}}} \\
      \rho_e\,\sigma_s\,\sigma_{\hat{s}\mid {\mathbf{e}}} & \sigma_{\hat{s}\mid {\mathbf{e}}}^2
    \end{pmatrix},
    \end{IEEEeqnarray*}
    where the correlation coefficient $\rho_e$ is defined as
    \begin{IEEEeqnarray*}{rcl}
    \rho_e = \frac{\mathrm{Cov}[\,s,\;\hat{s}\mid {\mathbf{e}}\,]}{\sigma_s \sigma_{\hat{s}\mid {\mathbf{e}}}}.
    \end{IEEEeqnarray*}
\end{proposition}
\begin{IEEEproof}
    We have shown that both $s$ and $\hat{s}\mid {\mathbf{e}}$ are asymptotically Gaussian. To demonstrate that they are jointly Gaussian, consider their expressions as linear transformations of the Gaussian vector $\mathbf{v}_q$:
    \vspace{-4pt}
    \begin{IEEEeqnarray*}{rcl}
    s = \mathbf{a}^T \mathbf{v}_q + \sum_{i=1}^{N}{c_i}, \quad \hat{s}\mid {\mathbf{e}} = \mathbf{a}^T \mathbf{D}_e \mathbf{v}_q + \sum_{i=1}^{N}{c_i},
    \end{IEEEeqnarray*}
    where \(\mathbf{D}_e = \mathrm{diag}(\mathbf{e})\). Stacking these equations, we obtain:
    \begin{IEEEeqnarray*}{rcl}
        \begin{pmatrix}
          s \\[6pt]
          \hat{s}\mid {\mathbf{e}}
        \end{pmatrix}
        =
        \underbrace{
          \begin{pmatrix}
            \mathbf{a}^T \\[3pt]
            \mathbf{a}^T \mathbf{D}_e
          \end{pmatrix}
        }_{A}
        \mathbf{v}_q
        \;+\;
        \underbrace{
            \sum_{i=1}^{N}{c_i} 
        }_{b}.
    \end{IEEEeqnarray*}
    Since $\mathbf{v}_q$ is asymptotically Gaussian with mean $\boldsymbol{\mu}$ and covariance $\boldsymbol{\Sigma}$, any linear transformation of $\mathbf{v}_q$ remains Gaussian. Therefore, the vector $\bigl(s,\,\hat{s}\mid {\mathbf{e}}\bigr)^T$ is jointly Gaussian with mean vector $A\,\boldsymbol{\mu} + b$ and covariance matrix $A\,\boldsymbol{\Sigma}\,A^T$.
    Specifically, the covariance between $s$ and $\hat{s}\mid {\mathbf{e}}$ is
    \vspace{-3pt}
    \begin{IEEEeqnarray*}{rcl}
    \mathrm{Cov}[\,s,\;\hat{s}\mid {\mathbf{e}}\,] = \mathbf{a}^T\,\mathbf{D}_e\,\boldsymbol{\Sigma}\,\mathbf{a}.
    \vspace{-2ex}
    \end{IEEEeqnarray*}
\end{IEEEproof}
\begin{theorem}
Given similarity scores $s$ and $\hat{s} \mid \mathbf{e}$, with means $\mu_s$, $\mu_{\hat{s} \mid \mathbf{e}}$, variances $\sigma_s^2$, $\sigma_{\hat{s} \mid \mathbf{e}}^2$, and correlation coefficient $\rho_e$, the probability of error $\mathbb{P}[s \hat{s} < 0 ]$ is given by:
\vspace{-3pt}
\begin{IEEEeqnarray*}{rcl}
\mathbb{P}\bigl[s\,\hat{s} < 0\bigr] = \sum_{\mathbf{e}} \big[\Phi(\delta) + \Phi(\hat{\delta}) - 2 \Phi_2(\delta, \hat{\delta}; \rho_e)\bigr]\; \mathbb{P}\bigl[\mathbf{e}\bigr].
\end{IEEEeqnarray*}
where $\delta = -\frac{\mu_s}{\sigma_s}$ and $\hat{\delta} = -\frac{\mu_{\hat{s} \mid \mathbf{e}}}{\sigma_{\hat{s} \mid \mathbf{e}}}$
Here, $\Phi(x)$ denotes the standard normal cumulative distribution function (CDF), and $\Phi_2(x, y; \rho_e)$ is the bivariate normal CDF with correlation coefficient $\rho_e$. The overall probability of error is obtained by averaging over all possible erasure patterns.
\end{theorem}
\begin{IEEEproof}
We begin by noting that the condition $s \hat{s} < 0$ translates to the events where $s$ and $\hat{s} \mid \mathbf{e}$ have opposite signs. This can be expressed as:
\begin{IEEEeqnarray*}{rcl}
\mathbb{P}[s \hat{s} < 0 \mid \mathbf{e}] = \mathbb{P}[s > 0, \hat{s} < 0 \mid \mathbf{e}] + \mathbb{P}[s < 0, \hat{s} > 0 \mid \mathbf{e}].
\end{IEEEeqnarray*}
To compute each term, we standardize $s$ and $\hat{s} \mid \mathbf{e}$ as:
\vspace{-3pt}
\begin{IEEEeqnarray*}{rcl}
Z_1 = \frac{s - \mu_s}{\sigma_s}, \quad Z_2 = \frac{\hat{s} \mid \mathbf{e} - \mu_{\hat{s} \mid \mathbf{e}}}{\sigma_{\hat{s} \mid \mathbf{e}}},
\end{IEEEeqnarray*}
\vspace{-3pt}
where $Z_1$ and $Z_2$ are bivariate normal random variables with zero mean, unit variances, and correlation coefficient $\rho_e$. 
Using this standardization, the conditions for $s > 0, \hat{s} < 0$ and $s < 0, \hat{s} > 0$ translate to:
\begin{IEEEeqnarray*}{rcl}
\mathbb{P}[s > 0, \hat{s} < 0 \mid \mathbf{e}] = \mathbb{P}[Z_1 > \delta, Z_2 < \hat{\delta}],
\end{IEEEeqnarray*}
\vspace{-20pt}
\begin{IEEEeqnarray*}{rcl}
\mathbb{P}[s < 0, \hat{s} > 0 \mid \mathbf{e}] = \mathbb{P}[Z_1 < \delta, Z_2 > \hat{\delta}],
\end{IEEEeqnarray*}
Using the properties of the bivariate normal distribution, we compute the probabilities as:
\begin{IEEEeqnarray*}{rcl}
\mathbb{P}[Z_1 > \delta, Z_2 < \hat{\delta}] = \Phi(\hat{\delta}) - \Phi_2(\delta, \hat{\delta}; \rho_e),
\end{IEEEeqnarray*}
\vspace{-20pt}
\begin{IEEEeqnarray*}{rcl}
\mathbb{P}[Z_1 < \delta, Z_2 > \hat{\delta}] = \Phi(\delta) - \Phi_2(\delta, \hat{\delta}; \rho_e).
\end{IEEEeqnarray*}
Adding these two terms, we have:
\vspace{-3pt}
\begin{IEEEeqnarray*}{rcl}
\mathbb{P}[s \hat{s} < 0 \mid \mathbf{e}] = \Phi(\delta) + \Phi(\hat{\delta}) - 2 \Phi_2(\delta, \hat{\delta}; \rho_e).
\end{IEEEeqnarray*}
Finally, to compute the overall probability of error, we take the expectation over all possible erasure patterns $\mathbf{e}$:
\begin{IEEEeqnarray*}{rcl}
\mathbb{P}[s \hat{s} < 0] = \mathbb{E}_{\mathbf{e}} \bigl[ P[s \hat{s} < 0 \mid \mathbf{e}] \bigr].
\end{IEEEeqnarray*}
This concludes the proof.
\end{IEEEproof}
{\color{blue}}
\subsection{Numerical Results}
In this section, we evaluate the performance of our proposed remote document retrieval system.
We evaluate our system by generating $n=2$ documents, each of length 10000, where the terms are sampled independently from a Zipf distribution considering a vocabulary size of $N=49000$.
Queries consist of $l=50$ terms, also
we excluded the top $l_s=10$ most frequent terms.
We vary the code rate $R$ and erasure probability $\epsilon$, examining their impact on retrieval error probability through 1000 Monte Carlo simulations. 

\begin{figure}[t]
    \centering
    \includegraphics[width=0.7\textwidth]{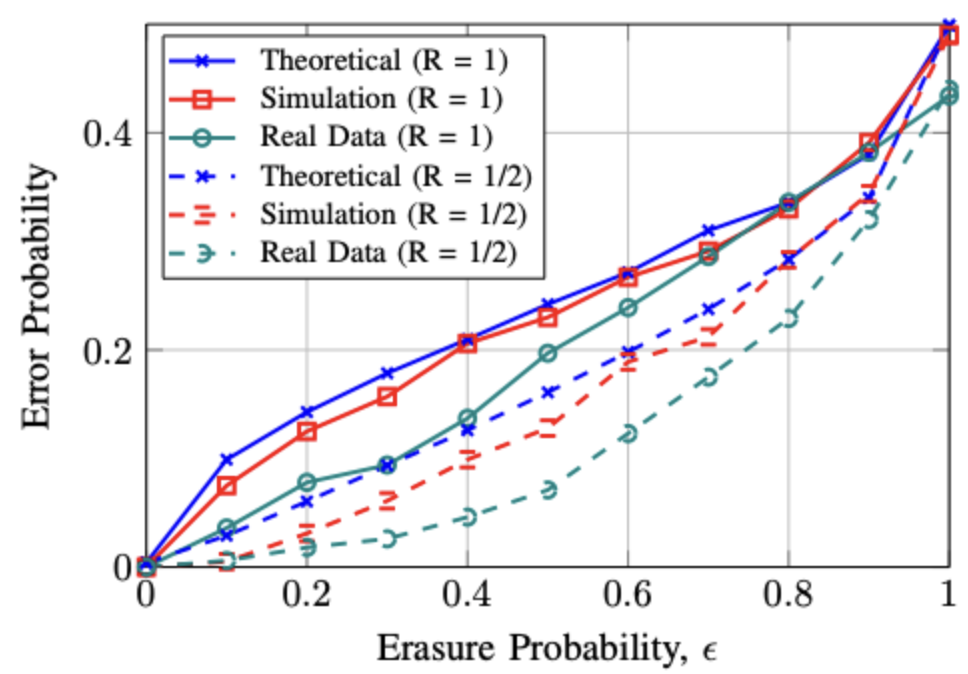}
    \caption{Analytical and numerical error probability results for the  L2-norm similarity measure versus query feature erasure probability under different coderates.}
    \label{fig:l2}
\end{figure}

\cref{fig:l2} illustrates the probability of error versus erasure probability $\epsilon$ for two scenarios, no repetition ($R=1$) and with repetition ($R=\frac{1}{2}$), for both analytical and numerical evaluations.
Note that for  the simulations the number of repetitions was determined using the rate formula in \cref{eq:repetition}, whereas the analytical results assume a fixed rate. 
However, since the query terms are sampled from Zipf's law, almost all important terms appear only once, resulting in roughly equal importance across all terms.
This makes the theoretical and simulation results comparable with respect to the effective coderate for each  term.

The close match between the simulation and theoretical results depicted in \cref{fig:l2} demonstrates the validity of our asymptotic analysis.
Although the theoretical results are derived under asymptotic assumptions, i.e., a large vocabulary and query lengths, they closely approximate the outcomes of finite-length simulations. This shows the robustness of the analytical framework in Theorem 1 and the validity of the Gaussian assumption for the similarity scores.
Additionally, the figure clearly illustrates the advantage of using repetition ($R=\frac{1}{2}$), as the error probability is consistently lower compared to the no-repetition case ($R=1$). 
This improvement highlights the advantage of assigning redundancy to important terms, which enhances resilience to erasures and significantly improves the overall retrieval performance.

{\color{black}To demonstrate our scheme’s performance on real-world data, we use the Google Natural Questions (NQ) dataset \cite{Kwiatkowski2019Nq} of one-query–one-answer pairs based on Wikipedia. We first select two long‑answer passages and use Llama‑3 \cite{Grattafiori2024Llama3} to generate 29 additional queries for each.
Then, we form a vocabulary from the unique terms in the documents and their associated queries.
Finally, we run simulations at rates $R=1$ and $R=\frac{1}{2}$ (see  \cref{fig:l2}).
As seen from this figure, while the curves obtained from the real dataset follow the same overall trend as both the theoretical and simulation results for the synthetic data, they consistently lie below the theoretical values, yielding lower error probabilities across the entire erasure range for both $R=1$ and $R=\frac{1}{2}$.
}

\bibliographystyle{IEEEtran}
\bibliography{main}

\end{document}